# Impacts of the Madden–Julian Oscillation on Precipitation Extremes in Indonesia

## Fadhlil R. Muhammad[1], Sandro W. Lubis[2], Sonni Setiawan[3]

[1]School of Earth Sciences, University of Melbourne, Australia

[2]Rice University, Houston, Texas, USA

[3]Department of Geophysics and Meteorology, IPB University, Indonesia

Correspondence

Fadhlil R. Muhammad, School of Earth Sciences, The University of Melbourne, Melbourne, Victoria, 3010, Australia

Email: fadhlilrizki@student.unimelb.edu.au

Funding information

Lembaga Pengelola Dana Pendidikan Indonesia, Beasiswa Ikatan Alumni Fakultas Ekonomi dan Bisnis Universitas Indonesia, and PPA Scholarship, DIKTI, Indonesia

The influence of the Madden-Julian Oscillation (MJO) on the precipitation extremes in Indonesia during the rainy season (October-April) has been evaluated using the daily station rain gauge data and the gridded Asian Precipitation–Highly Resolved Observational Data Integration Toward Evaluation of Water Resources (APHRODITE) from 1987 to 2017 for different phases of the MJO. The results show that MJO significantly modulates the frequency of extreme precipitation events in Indonesia, with the magnitude of the impact varying across regions. Specifically, the convectively active (suppressed) MJO increases (decreases) the probability of extreme precipitation events over the western and central parts of Indonesia by up to 70% (40%). In the eastern part of Indonesia, MJO increases (decreases) extreme precipitation probability by up to 50% (40%). We attribute the differences in the probability of extreme precipitation events to the changes in the horizontal moisture flux convergence induced by MJO. The results indicate that the MJO provides the source of predictability of daily extreme precipitation in Indonesia.

## 1 | INTRODUCTION

Madden-Julian Oscillation (MJO) is the major intraseasonal (30-60 days) variability in the tropics, characterized by the eastward propagation of cloud clusters and precipitation from the Indian Ocean to western Pacific Ocean (Madden and Julian, 1971). The peak MJO signals are located over the Indian Ocean (eastern Pacific Ocean) during boreal winter (boreal summer) and are dominated by mean westerlies or weak mean zonal winds at 850 hPa and the surface (Zhang and Dong, 2004). During boreal winter, the interaction of MJO and topography causes the MJO to propagate eastward and southward in the Maritime Continent (MC) before it reaches the western Pacific Ocean (Inness and Slingo, 2006; Peatman et al., 2013). MJO brings in organized convection and associated circulation favouring a region for rainy or dry conditions depending on its phase. Previous studies have shown that MJO can influence the frequency and intensity of global precipitation and temperature events (Barlow et al., 2005; Barlow and Salstein, 2006; Jeong et al., 2008; Lin et al., 2010; Lu et al., 2012; Ren and Ren, 2017; Seto et al., 2004; Wheeler et al., 2009; Zhou et al., 2012).

Precipitation in Indonesia exhibits substantial variability at the synoptic to intraseasonal time scales (Aldrian et al., 2004; Aldrian and Dwi Susanto, 2003; As-syakur et al., 2013; Harry H. Hendon, 2003; Kripalani and Kulkarni, 1997; Moron et al., 2010; Qian et al., 2010; Rajagopalan et al., 2016). On the interannual time-scales, El-Nino Southern Oscillation (Aldrian and Dwi Susanto, 2003; Hamada et al., 2012; Harry H Hendon, 2003; Rakhman et al., 2017) and Indian Ocean Dipole (Hamada et al., 2012; Muhammad et al., 2019; Nur'utami and Hidayat, 2016) are the largest contributors to monthly precipitation variance, while the southeast and northwest monsoons strongly influence seasonal precipitation variance in Indonesia (Aldrian and Dwi Susanto, 2003; Hamada et al., 2002). The monsoonal precipitation is mostly observed over the Indonesian regions with a rainy (dry) season starting from November to February (June to August) (Aldrian and Dwi Susanto, 2003). On the daily to intraseasonal time-scales, the variability of precipitation in Indonesia is explained by MJO and convectively coupled equatorial waves (CCEWs). MJO that contributes up to 50% of the total variance (Waliser et al., 2009), while CCEWs contribute up to 12% of the total variance (Lubis and Jacobi, 2015).

The influence of the MJO on Indonesian precipitation variability has been generally discussed by several authors (Jones et al., 2004; Hidayat and Kizu, 2010; Xavier et al., 2014). A detailed explanation of the influences of MJO on precipitation in Indonesia has been demonstrated by Hidayat and Kizu (2010). They showed that during austral summer, the MJO increases (decreases) Indonesian precipitation by up to 5 mm/day during phases 2 to 4 (phases 6 to 8). In case of precipitation extremes, the convectively active phases of MJO increase extreme precipitation events in the tropical precipitation as part of a global response by approximately 15-20% (Jones et al., 2004). More recently, Xavier et al., (2014) quantified the probability changes in the extreme precipitation over Southeast Asia and its relationship to large-scale circulation during boreal winter. They found that MJO increases (decreases) the probability of extreme precipitation by up to 30-50% on phases 2 to 4 (10-20% on phases 6 to 8). While Xavier et al., (2014) have provided a general view on the MJO impacts on extreme precipitation over Southeast Asia, a detailed analysis of such links on the spatial distribution of "regional" precipitation extremes over the land regions Indonesia remains unclear and is worthy for further investigation. Understanding the impact of MJO on regional precipitation extremes in Indonesia is essential, owing to the different nature of MJO characteristics over different parts of Indonesia (Hidayat and Kizu, 2010). The two main questions we would like to address in this study are as follow:

1.   What are the effects of MJO on regional precipitation extremes in Indonesia?

2.   What is the underlying dynamics of the MJO impacts on regional precipitation extreme events in Indonesia?

The data sets and methods are described in Section 2. Results and discussions are presented in Section 3. The conclusions of the study are offered in Section 4.

## 2  |  DATA AND METHODS

To study the impact of the MJO on extreme daily precipitation in Indonesia, we use gridded rainfall data from Asian Precipitation - Highly Resolved Observational Data Integration Towards Evaluation of Water Resources (APHRODITE) (V1901, Yatagai et al., 2012) from 1998 to 2015, with a spatial resolution of 0.25 x 0. 25°. APHRODITE comprises of data collected from 12000 rain-gauge stations across the globe. The interpolation was done by taking the topography effects into account, resulting in a more accurate representation of the precipitation over the mountainous regions (Schaake, 2004). Furthermore, in order to support the results obtained from APHRODITE, we also use extensive high-quality rain-gauge database operated by BMKG for the period of 1987 to 2017, from 63 stations in Indonesia (Fig. 3b). We only consider the data that have less than 20% of missing observations.

To understand the dynamical links between extreme precipitation events and large-scale circulations, we use interpolated outgoing longwave radiation (OLR) as a proxy for convection obtained from NOAA/NESDIS (Liebmann and Smith, 1996). Furthermore, we also use the relative humidity, horizontal and vertical wind, and air temperature data at various levels (1000-100 hPa) obtained from the National Center for Environmental Prediction R-2 (NCEP/DOE Reanalysis 2) (Kanamitsu et al., 2002) with a spatial resolution of 2.5 x 2.5°.

The MJO events are defined based on Real-time Multivariate MJO Index (RMM) (Wheeler and Hendon, 2004). The multivariate EOF is calculated from the zonal wind anomalies at 850 and 200 hPa, and OLR anomalies. The two leading principal components of the multivariate EOF are then defined as the real multivariate MJO indices (RMM1 and RMM2). These indices are then used to calculate the amplitude of MJO events as $\sqrt{(RMM1^2 + RMM2^2)}$. The MJO is considered as a strong (weak) event if the amplitude is higher (less) than 1. Furthermore, the MJO event is divided into eight phases, where each phase of MJO indicates the location of the MJO convective centre. The convective centre of MJO propagates eastward from west Africa (phase 1) to the east, passing over the Indian Ocean (phase 2 and 3), MC (phase 4 and 5), and decaying over the western Pacific (phase 7 and 8).

In order to examine the probability changes of extreme precipitation, we use a probability composite analysis as described in Wheeler et al. (2009). First, we select the data for the period of the rainy season (from October to April) and then separate them into "strong" MJO events for each phase, as in Wheeler and Hendon (2004). Second, we calculate the cumulative probability of daily precipitation events that exceed the $95^{th}$ percentile of precipitation for all days in the season ($P_{ALLDAYS}$) and each phase of MJO ($P_{MJO}$). Finally, we calculate the probability changes as (Ren and Ren 2017):

$$\Delta P = \frac{(P_{MJO} - P_{ALLDAYS})}{P_{ALLDAYS}} \times 100\%, \tag{1}$$

where $\Delta P$ is the probability changes in extreme events for each phase of MJO.

To elucidate the underlying dynamics of the MJO's impact on precipitation extremes, we also analyze the composites of OLR, vertical wind component ($\omega$), vertically integrated moisture flux convergence (*VIMFC*), and vertical moisture advection anomalies for each phase of MJO. The moisture flux convergence and vertical moisture advection are derived from the moisture budget equation defined as (see Banacos and Schultz, 2005):

$$-\underbrace{\left(\frac{\partial q}{\partial t}\right)'}_{\substack{local \\ moisture \\ tendency}} - \underbrace{(\overrightarrow{\nabla q} \cdot (q\,\boldsymbol{u_H})\,)'}_{\substack{horizontal \\ moisture\,flux \\ convergence}} - \underbrace{\left(q\,\frac{\partial \omega}{\partial p}\right)'}_{\substack{vertical \\ moisture \\ convergence}} - \underbrace{\left(\omega\,\frac{\partial q}{\partial p}\right)'}_{\substack{vertical \\ moisture \\ advection}} = \underbrace{(P_r - E)'}_{\substack{sources\,and\,sinks}}, \tag{2}$$

where $q$ is specific humidity, $p$ is pressure, $\boldsymbol{u_H}$ is horizontal components of wind, $\omega$ is a vertical component of wind, $E$ is evaporation, and $P_R$ is precipitation. The primes denote filtered anomaly fields with a high-frequency cut-off of 20 days and a low frequency of 100 days to retain the MJO signals. In order to obtain the *VIMFC*, the horizontal moisture flux convergence term is then vertically integrated as follows (van Zomeren and van Delden, 2007):

$$VIMFC = -\frac{1}{g}\int_{1000\,hPa}^{100\,hPa}\left[-(\overrightarrow{\nabla}\cdot(q\,\boldsymbol{u_H})\,)'\right]dp, \tag{3}$$

where $g$ is gravity (9.81 kg/ms$^2$). Due to the lack of specific humidity data in NCEP/DOE Reanalysis 2, we calculate the specific humidity from this formulation:

$$q = q_s \times RH, \tag{4}$$

where $RH$ is relative humidity and $q_s$ is saturated specific humidity calculated as:

$$q_s = \frac{(0.622 \times e_s)}{(p - 0.378 \times e_s)}, \tag{5}$$

where $p$ is pressure and $e_s$ is saturated pressure, which can be calculated using the Clausius-Clapeyron formula as:

$$e_s = 6.11\,mb \times \left[\frac{l_{w,v}}{R_v} \times \left(\frac{1}{273.15} - \frac{1}{T}\right)\right], \tag{6}$$

$l_{w,v} = 2.5 \times 106$ Jkg$^{-1}$ is the latent heat of transformation of water to vapour, $R_v = 461.5$ Jkg$^{-1}$K$^{-1}$ is specific gas constant for water vapour, and $T$ is temperature.

The significant test for each composite is done by using a bootstrap method (Wilks, 2005). In this method, for each phase of the MJO, we generate 1000 synthetic composites by randomly selecting sample of events to derive a bootstrapped mean and confidence limits. The 1000 synthetic composites are then sorted to find the 2.5$^{th}$ and 97.5$^{th}$ percentiles for a two-tailed test, with significance at 95% level. Then, we compare the real composites for each phase of MJO with the percentile levels of the synthetic composites.

## 3 | RESULTS AND DISCUSSIONS

## 3.1 | MJO Impact on Extreme Precipitation in Indonesia

### 3.1.1 | Spatial Distribution of Precipitation Anomalies

We begin our analysis with an examination of the MJO impact on the spatial distribution of precipitation anomalies over Indonesia. Figure 1 shows the composite of APHRODITE precipitation anomalies (in mm/day) at each MJO phase during the rainy season. In general, the increase in precipitation tend to reach their maxima over the large landmasses of Indonesia during phases 2 to 4 (Fig. 1(b)-(d)). In phases 5 to 8 (Fig. 2(e)-(h)), the variation of precipitation anomalies exhibits a pattern similar to those in phases 1 to 4 (Fig. 1(a)-(d)), but with the opposite sign. The average increases (decreases) of precipitation in phases 2 to 4 compared to phases 5 to 8 is about 2 - 6 mm/day, approximately 20 – 30% of the seasonal mean precipitation (Fig. 3a). In addition, it is clearly seen that the positive precipitation anomalies due to MJO has already been observed in Indonesia during phases 1 to 2, while the MJO convective centre is still located over the Indian Ocean. This leaping ahead of the MJO envelope is known as a "vanguard" effect of precipitation (Matthews et al., 2013; Peatman et al., 2013). This effect is seen only over the MC and is mainly triggered by the diurnal cycle of convective activity ahead of the MJO envelope and by the enhanced topographic frictional moisture convergence associated with the Kelvin and Rossby waves (Matthews et al., 2013; Peatman et al., 2013).

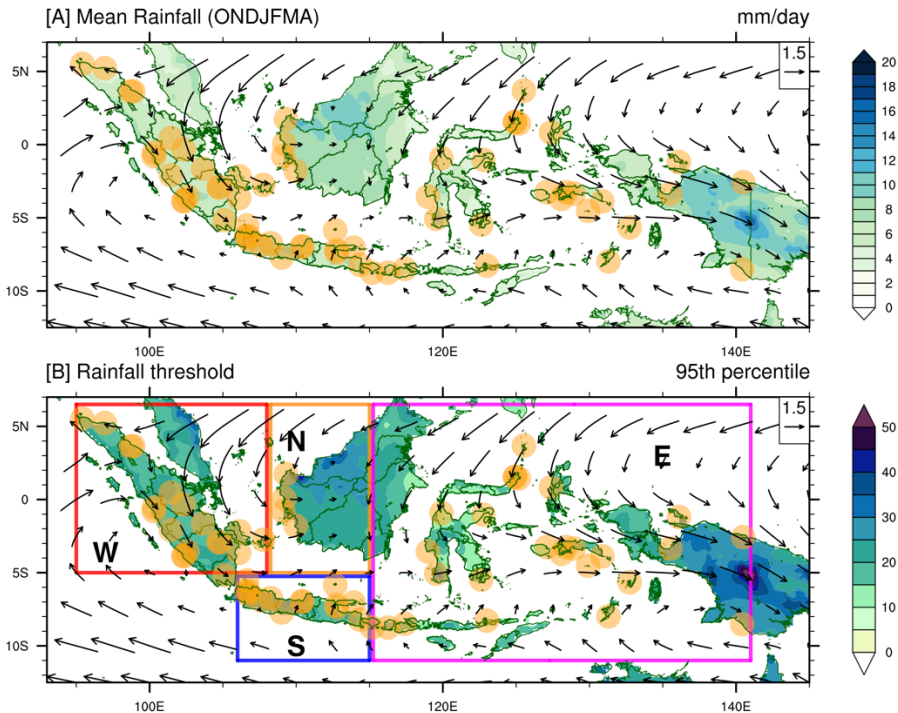

F I G U R E  1.  Composites of precipitation anomalies in each MJO phase (a – h) during the rainy season (October-April) in Indonesia based on APHRODITE. Dots indicate values significant at 95% confidence level.

The variation of the MJO-induced precipitation anomalies obtained from the APHRODITE (Fig. 1) is consistent with the observed precipitation data from the rain-gauge stations (Fig. 2). In particular, the MJO tends to increase (decrease) the precipitation during phases 2 to 4 (phases 5 to 8), by which the average increase

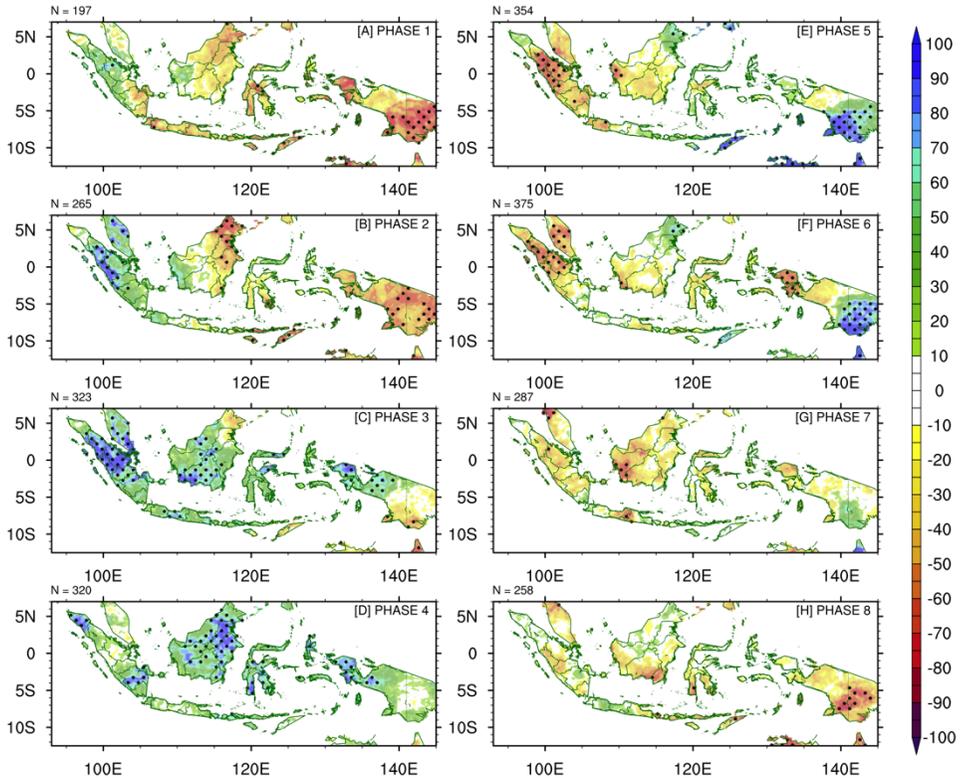

F I G U R E 2. Composites of precipitation anomalies observed by rain gauges (triangle) and the APHRODITE (shading). Filled triangle and dots indicate values significant at 95% confidence for the rain gauge data and the APHRODITE data, respectively.

(decrease) of precipitation anomalies is about 1 - 4 mm/day. In addition, the precipitation response from the rain-gauge stations is more inhomogeneous compared to that of the APHRODITE. This inhomogeneous response could be due to strong local effects in some regions in Indonesia, such as over the western and central part of Java (Hidayat and Kizu, 2010).

In summation, the results show that the influence of MJO on the precipitation over Indonesia is robust in both the rain-gauge data and gridded rainfall data, especially during phases 2 to 4 (phases 6 to 8) for the positive (negative) anomalies. Next, we will examine how the MJO influence the likelihood of extreme precipitation events in Indonesia.

## 3.1.2 | Spatial Distribution of Precipitation Extremes

Figure 3 shows the distribution of mean and 95th percentile precipitation at each grid point during the rainy season (October – April). The 95$^{th}$ percentile precipitation is used

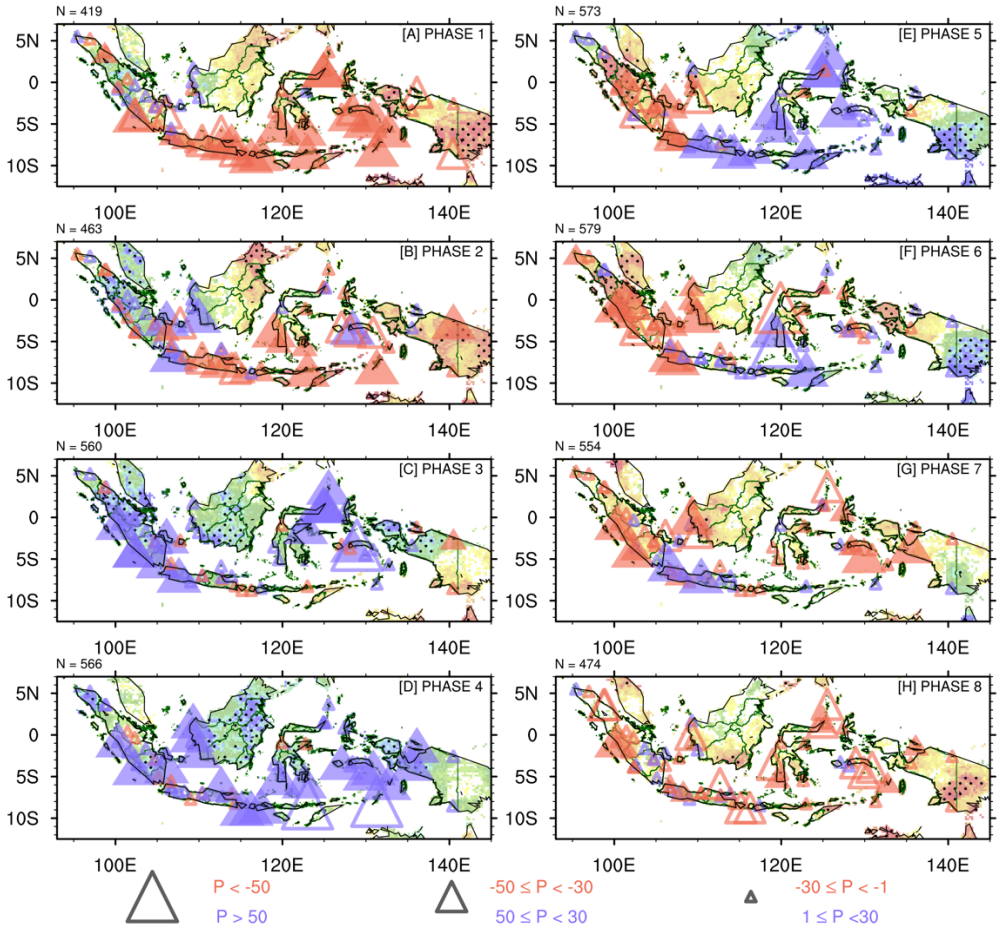

F I G U R E 3. (a) Mean and (b) 95th percentile of daily precipitation superimposed with 850 hPa horizontal wind during the rainy season (October-April). Circles denote the location of in situ measurements of precipitation.

as a threshold to determine extreme precipitation events (Wheeler et al., 2009; Xavier et al., 2014). We quantify the impact of MJO on extreme precipitation as the percentage change of the cumulative probability of extreme events ($P_{MJO}$) relative to the baseline probability ($P_{ALLDAYS}$) (see Eq. 1).

Figure 4 shows the percentage of changes in extreme precipitation probability for each MJO phase from the observed gridded rainfall data. In general, the changes in the probability of extreme events due to MJO are consistent with the distribution of MJO-induced precipitation anomalies (Fig 1 and 2), in which the phases 2 to 4 (phases 6 to 8) produce significantly increased (decreased) extreme precipitation probabilities over the region (see Fig. 4 and 5). A cursory inspection of Figure 4 shows that the increase in the probability of extreme events over the land regions on phase 2 is up to 60% over the west of Sumatra and Borneo during the rainy season (Fig. 4b). The maximum increase is observed during phase 3 (Fig. 4(c)-(d)) by up to 100% over the western coast of Sumatra and 70-80% over Java and Papua (Fig. 4c). On the other hand, the increase in the probability during phase 4 is higher over Borneo and Sulawesi compared to that of phase 3. However, the increase is significantly lower over Sumatra and Java (Fig. 4c and 4d). During phase 5, the MJO decreases the probability of extreme events over most of the western part of Indonesia (e.g. Sumatra and west of Borneo), while increases the probability over the southeastern part of Indonesia (Fig. 4e). Finally, as the suppression phases of the MJO becomes dominant during phases 6 to 8 (Fig. 4(f)-(h)), the MJO mainly decreases the probability of extreme events by up to 80% over Sumatra, Borneo, Java, Sulawesi, and Papua.

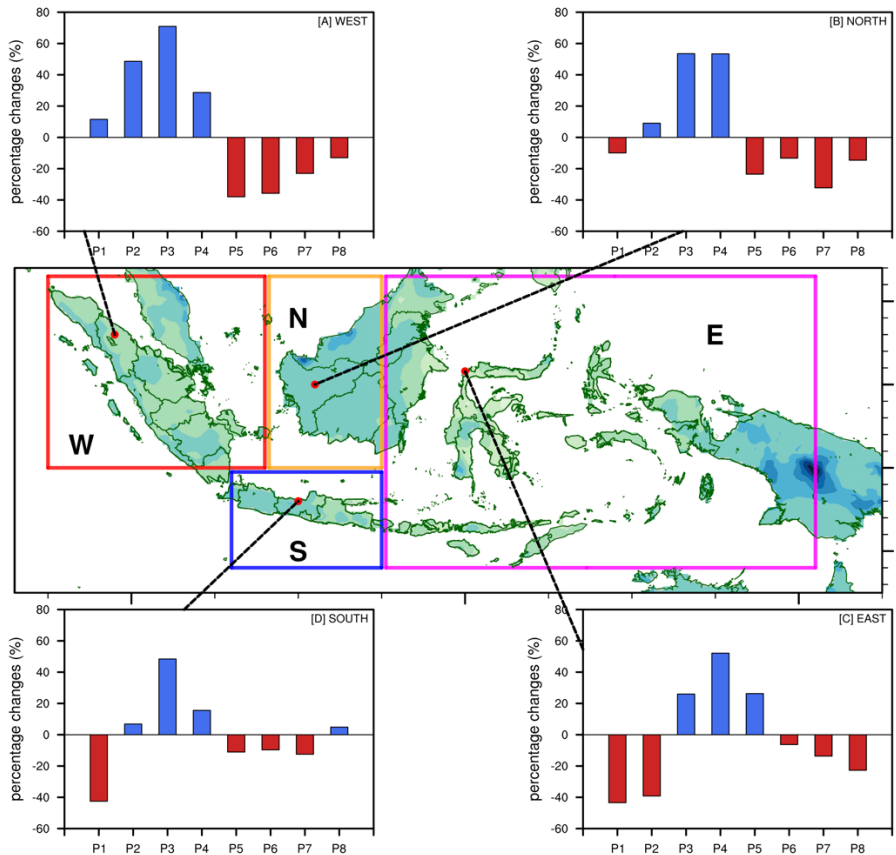

F I G U R E  4. Percentage changes in the probability of extreme events during different phases of MJO (a - h). Dots indicate values significant at 95% confidence level.

In order to validate the results obtained from APHRODITE over the land regions, we also examine the impact of MJO on extreme precipitation events based on rain-gauge data. Figure 5 shows the probability composites of extreme events observed by rain-gauge stations for each phase of the MJO. We find that the impact of MJO on extreme events observed by rain-gauge stations is generally consistent with the APHRODITE (Fig. 4). The increase (decrease) in the probability of extreme events is evident at phases 2 to 4 (phases 6 to 8) (Fig. 5(b)-(d) and Fig. 5(f)-(h)). The strongest impact is observed during phases 3 to 4, which increases the probability of extreme precipitation events by more than 50% (Fig. 5(c)-(d)). During phases 6 to 8, MJO mainly decreases the probability of extreme precipitation events in Indonesia (Fig. 5(f)-(h)), except over the western and central parts of Java during phase 7 or 8 that could be affected by the local effects, such as topography (Hidayat and Kizu, 2010) and the effect of diurnal cycle (Peatman et al., 2014).

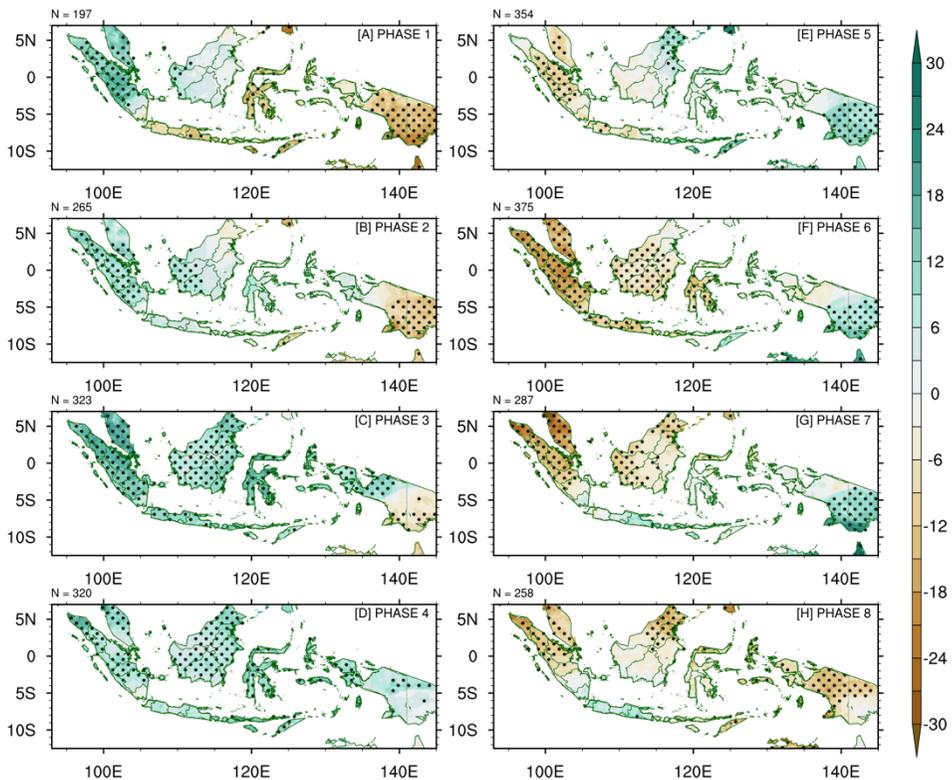

FIGURE 5. Percentage changes in the probability of extreme events observed by rain gauges (triangle) and the APHRODITE (shading). Filled triangle and dots indicate values significant at 95% confidence for the rain gauge data and the APHRODITE data, respectively.

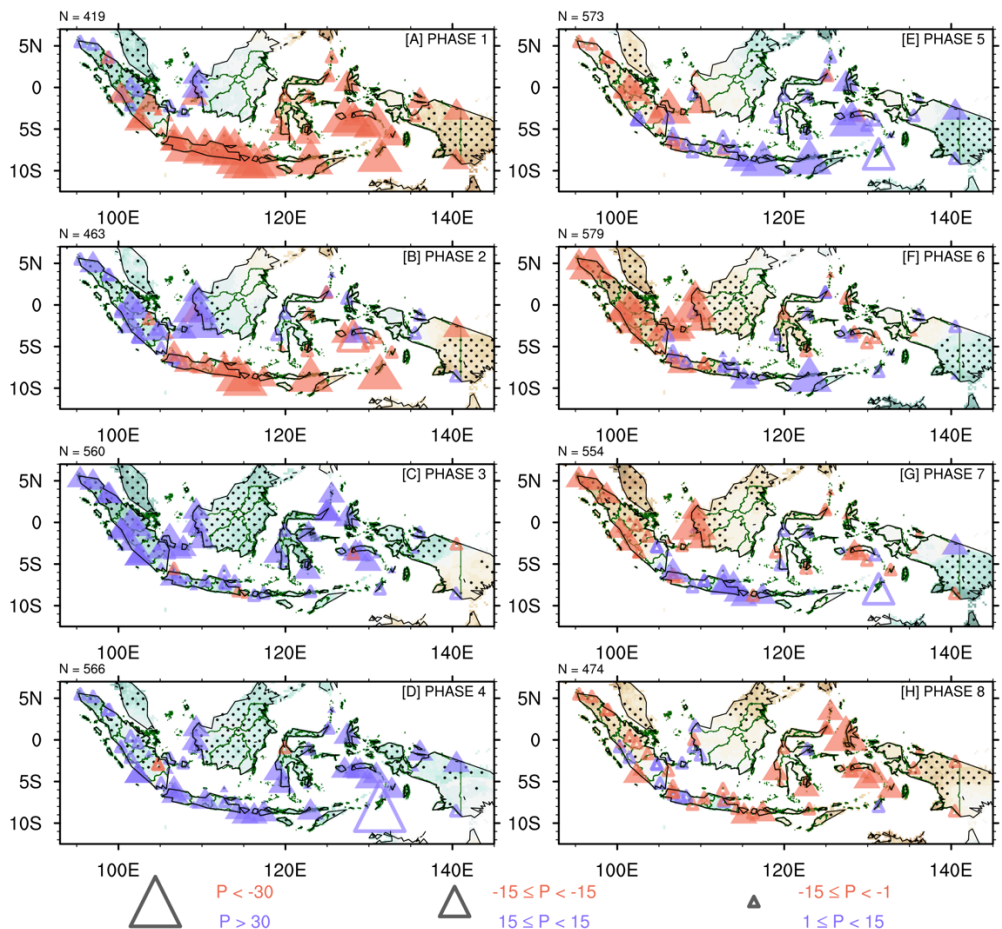

F I G U R E 6. Percentage changes in the probability of extreme events based on APHRODITE data over the region of interest during different phases of MJO. The thresholds (95th percentile) are drawn on the map.

To better understand the impact of MJO on regional extreme precipitation events, we further calculate average percentage of changes in probability for four sectors in Indonesia, including, the West, North, South, and eastern parts of Indonesia based on their geographical locations (Fig. 6). In general, it can be clearly seen that the impact of MJO is the strongest over the western part of Indonesia, with a maximum probability of around 70% or about 20% higher than the other regions. In particular, the increase (decrease) in extreme probability over the western part of Indonesia begins from phases 1 to 4 (5 to 1), with the maximum increase (decrease) up to 70% (40%) during phase 3 (phase 5) (Fig. 6). Furthermore, as the convective envelope of MJO propagates further eastward, the associated impact also migrates to the northern and southern parts of Indonesia. Both sectors experience an increase in the probability of extreme events during phases 2 to 4, with the maximum by up to 50% during phase 3. In the southern (northern) part of Indonesia, the maximum decrease occurs during phase 1 (phase 7) by up to 40% (35%). On the other hand, the increase (decrease) of extreme probability over the eastern part of Indonesia occurs from phases 3 to 5 (phases 6 to 2), with the maximum increase (decrease) by about 50% (40%) during phase 4 (phase 1).

Overall, the results indicate that the impact of MJO on extreme precipitation events is robust and varying across regions in Indonesia. The probability of extreme precipitation increases (decreases) on the days when the MJO wet (dry) phase is occurring. In the next section, we will investigate the underlying mechanisms that are responsible for elucidating such impacts.

## 3.2 | Dynamical Links between Precipitation Extremes and MJO

To understand the dynamical factors contributing to the MJO-induced extreme precipitation events, we examine two important processes that are responsible for modulating the precipitation anomalies, namely the vertical moisture flux convergence of moisture (van Zomeren and van Delden, 2007) and the vertical advection of moisture (Benedict and Randall, 2007).

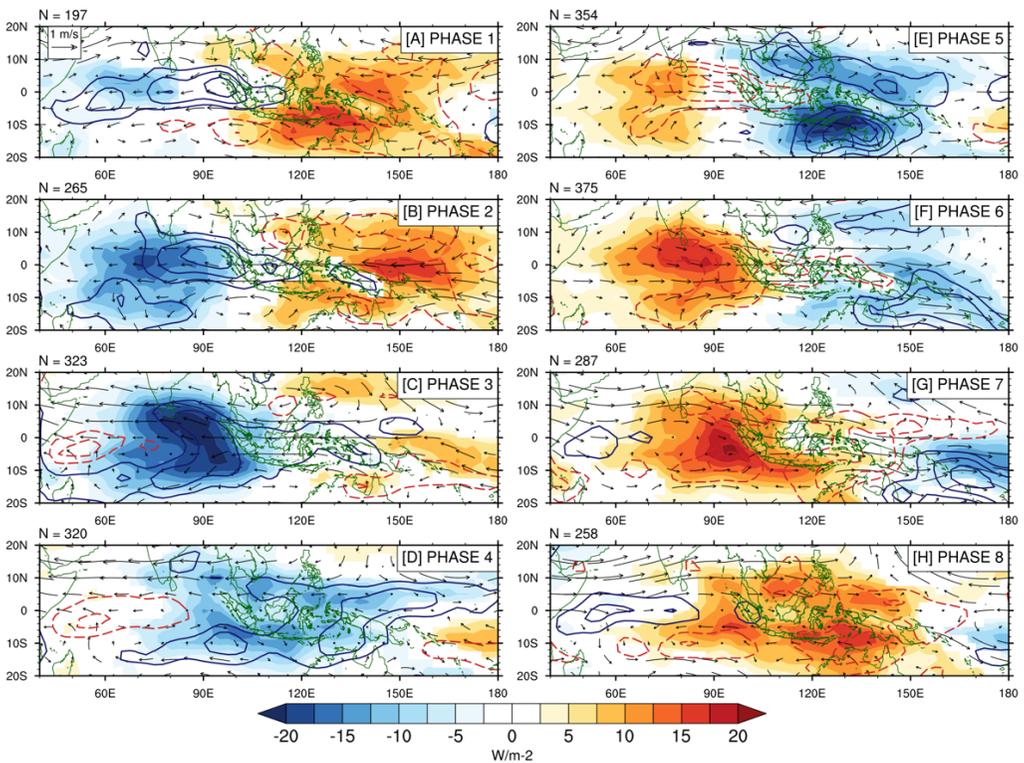

F I G U R E  7.  Composites of VIMFC anomalies (contour) (x 10-6 kg m-1s-1) and OLR (shading) (W m-2)  superimposed with horizontal wind (vector) (m s-1) during different phases of MJO. Contours are from -6 to 6 with 1 interval. Shaded values are significant

Figure 7 shows composite anomalies of OLR (shading) and VIMFC (contour) for each phase of the MJO during the rainy season. It is shown that the enhanced (suppressed) convective activity and moisture convergence (divergence) are linked to the positive (negative) changes in precipitation anomaly and extreme precipitation probability. During phases 1 and 2, the enhanced moisture convergence has already extended over the western and northern parts of Indonesia (Fig. 7(a)-(b)), although the enhanced convection associated with MJO has not reached these regions yet. This enhanced moisture convergence contributes

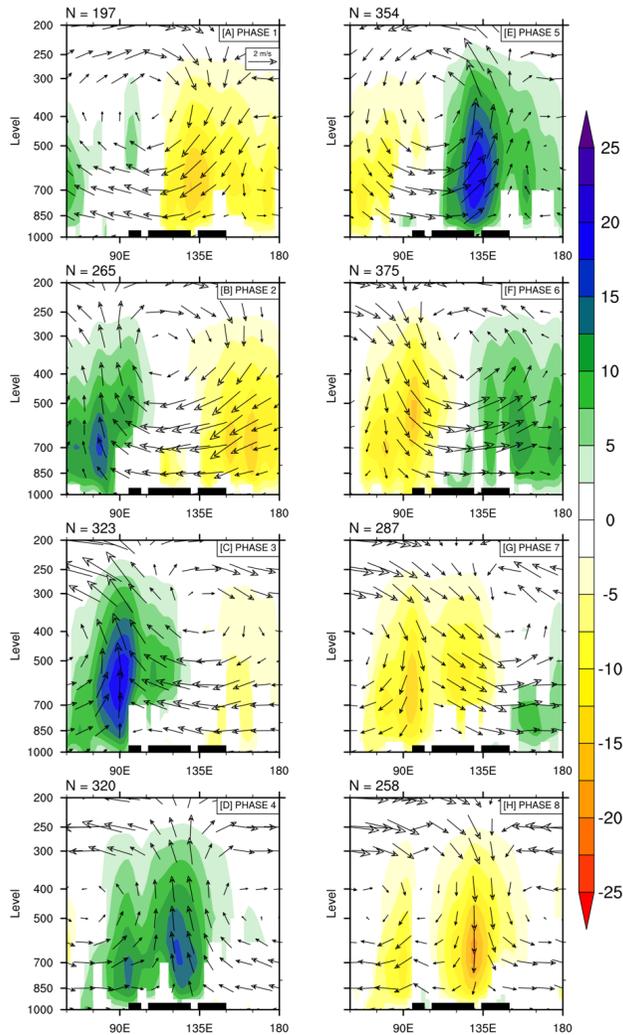

**F I G U R E  8.**  Vertical cross-section of vertical moisture advection (x10-6 s-1) (shading) and wind (vector) anomalies during different phases of MJO. Shaded values are significant at the 95% level.

to an increase in precipitation anomaly and extreme precipitation probability (Fig. 1 – 2 (a)-(b) and 4 - 6 (a)-(b)), which is likely to be associated with topography-enhanced Kelvin wave responses over those regions (Matthews et al., 2013; Peatman et al., 2013). Furthermore, we observed strong convergence over the western part of Indonesia during phase 3  (Fig. 7c), which is overall consistent with an increase in the precipitation anomaly and probability of extreme precipitation (Fig. 1 – 2 (c), 4 – 5 (c), and 6a). The enhanced moisture convergence is coupled with intense deep convection over this region, resulting in an increase in the probability of extreme events  (Fig. 4 – 5 (c), and 6a). As the convectively active phase of MJO moves eastward over to the MC during phase 4 (Fig. 7d), we observed moisture convergence extending over most regions in Indonesia, but with weaker convective activity compared to the phase 3. The moisture convergence and convective activity during phase 4 are consistent with enhanced precipitation and extremes over most regions in Indonesia. During phases 5 to 8 (Fig. 7(e)-(h)), the variation of VIMFC and OLR anomalies exhibits a pattern similar to that of phases 1 through 4 (Fig. 7(a)-(d)), but with the reversed sign. This indicates that the decrease in precipitation anomalies and extreme events are caused by a decreased horizontal moisture flux convergence induced by MJO (Fig. 1 - 2 and 4 - 5(e)-(h)).

The enhanced (suppressed) VIMFC and OLR in each phase of MJO are consistent with enhanced (suppressed) upward (downward) motion of moist air over the region. Figure 8 shows a cross-section of vertical moisture advection anomalies (color shading), and wind velocity (vectors) for each phase of MJO averaged over $11°S$-$11°N$. It can be clearly seen that an increase (decrease) of precipitation and extremes is linked to an increase of upward (downward) moisture advection in the troposphere during the rainy season. In particular, during phases 2 through 4 (Fig. 8(b)-(d)), the upward motion of moist air favors moisture entrainment at low to the middle-level troposphere. This supports the development of convection over the MC, consistent with previous findings (Benedict and Randall, 2007; Xavier et al., 2014). In contrast, the downward advection of moisture to the east of central convection of MJO depletes water vapor at the mid-level troposphere (Wang et al., 2019). During phase 3, we observe enhanced vertical moisture advection from $90°E$ to $120°E$ over the western parts of Indonesia (Fig. 8c). It is seen that the strongest advection is observed over the western part of Sumatra ($90$-$95°E$) between $1000$-$300$ hPa level., while a downward advection is observed between $1000$-$700$ hPa level over the eastern parts of Indonesia ($135$-$145°E$). This strong advection accompanied by the convergence of westerly wind burst and local easterly mountain wind over Sumatra (Fig. 7c and 8c) further induces strong MJO impact on the precipitation over Sumatra (Fig. 1-2(c) and 4-5(c)) (Wu et al., 2017). As the upward moisture advection moves eastward during phase 4, the increase in the precipitation and extreme events is observed over most regions in Indonesia (Fig. 1 - 2(d) and Fig.4 - 5(d)). During phase 5, we observe enhanced upward moisture advection over $120$-$140°E$ (Fig. 8e). This upward vertical motion is noticeably strong at $135°E$ and between $1000$-$300$ hPa, consistent with high precipitation over the eastern part of Indonesia (Fig. 4 - 5(e) and 6c). Finally, throughout phases 6 - 8 (Fig. 8(f)-(h)), a similar pattern is observed as in phases 2 to 4 (Fig. 8(b)-(d)), except with reversed signs. During these phases, the downward moisture advection occurred across most regions in Indonesia and is consistent with relatively drier days in these regions.

To sum up, we find that the enhanced (suppressed) moisture flux convergence induced by the MJO is the key mechanism that contributes to the increased (decreased) precipitation anomaly and extreme precipitation events over Indonesia during different phases of MJO. This enhanced (suppressed) VIMFC is consistent with enhanced (suppressed) upward (downward) motion of moist air over these regions.

## 4 | SUMMARY AND DISCUSSION

We have examined the impacts of the MJO on the frequency of precipitation extremes in Indonesia during the rainy season (October-April) using the high-quality daily rain gauge data from 63 stations and the gridded rainfall APHRODITE data. We found that MJO can significantly modulate the frequency and intensity of extreme precipitation events in Indonesia. A detailed analysis of the impacts of MJO on extreme precipitation in Indonesia reveals the following key results:

1. The convectively active phases of MJO increase the probability of extreme precipitation events by more than 50% over most areas in Indonesia during phases 3 to 4.

2. The convectively suppressed phases of MJO decrease the probability of extreme precipitation events by about 40–70% over Sulawesi and Papua during phases 1 to 2, and over Java, Sumatra, and Borneo islands during phases 5 to 6.

3. The western part of Indonesia experiences the strongest impact of the MJO, while the impact over the southern part of Indonesia is relatively weaker due to strong interference with local effects.

4. The impact of MJO on extreme precipitation over Indonesia can be explained through the changes in the moisture flux convergence and the vertical advection of the moist air. The increase (decrease) in MJO-induced moisture flux convergence leads to a high (low) likelihood of extreme precipitation in Indonesia.

It is noteworthy that the impacts of the MJO on extreme precipitation events are largely inhomogeneous over the land regions. It is possible that this is caused by a unique interaction between MJO-induced circulation and lower boundary forcing such as topography and local effects (Hsu and Lee, 2005; Wu and Hsu, 2009; Hidayat and Kizu, 2010; Kim et al., 2017). In addition, the role of other atmospheric variability with different timescales can also affect the extreme precipitation in some regions of Indonesia. For example, the quasi-biweekly oscillation is partly responsible for the increase of extreme precipitation probability over the western part of Sumatra (Wen and Zhang, 2008), while the influence of cold surge and Borneo vortex can modulate the extreme precipitation probability over the western and northern parts of Borneo (Chang et al., 2005; Lim et al., 2017). A detailed investigation of the interactions between MJO and topography and other atmospheric variability over the MC, as well as the underlying mechanisms of the inhomogeneous impact of MJO on precipitation extremes are subjects for a future study.

Overall, the results indicate that the MJO provides the source of predictability of extreme precipitation in Indonesia. This suggests that the skill in probabilistic prediction of extreme daily precipitation in Indonesia would be dependent on MJO prediction skill in the operational weather models and the modelled relationship between MJO and convection.

## 5  |  ACKNOWLEDGEMENTS

We are grateful for the generous and insightful comments of Matthew Wheeler, Givo Alsepan, Rahmat Hidayat, and Eddy Hermawan. This work was supported by Ikatan Alumni Fakultas Ekonomi dan Bisnis Universitas Indonesia (ILUNI FEB UI) Scholarship, Direktoran Jenderal Pendidikan Tinggi (DIKTI) under PPA Scholarship, and Lembaga Pengelola Dana Pendidikan (LPDP) Scholarship. More detailed information about the data can be found at the APHRODITE web page (http://aphrodite.st.hirosaki-u.ac.jp/products.html) and BMKG data online web page (http://dataonline.bmkg.go.id/home).

## 6  |  CONFLICT OF INTERESTS

The authors report that they have no conflict of interest.